# A simplified method characterizing magnetic ordering modulated photo-thermoelectric response in noncentrosymmetric semimetal Ca$_3$Ru$_2$O$_7$

*Qiang Chen[1], Jialin Li[1], Huanfeng Zhu[1], Tian Zhang[1], Wei Tang[1], Hui Xing[2], Jin Peng[3], Zhiqiang Mao[4], Linjun Li[1]\**

Qiang Chen, Jialin Li, Huanfeng Zhu, Tian Zhang, Wei Tang, Linjun Li

State Key Laboratory of Modern Optical Instrumentation, College of Optical Science and Engineering, Zhejiang University, Hangzhou 310027, P. R. China

Hui Xing

Key Laboratory of Artificial Structures and Quantum Control, and Shanghai Center for Complex Physics, School of Physics and Astronomy, Shanghai Jiao Tong University, Shanghai 200240, P. R. China

Jin Peng

School of Physics, Southeast University, Nanjing 211189, P. R. China

Zhiqiang Mao

Department of Physics, Pennsylvania State University, University Park, PA 16802, USA

E-mail: lilinjun@zju.edu.cn

**Photo-Thermoelectric (PTE) response is usually one of the main working mechanisms for photodetectors. However, as another fast and easier way to measure thermoelectric characteristics of materials, it can also reveal important physics such as electric-phonon coupling, electron-electron correlation, etc. Recently, the spin entropy related to magnetic order transition which contributes to thermoelectric power is attracting more and more attention. Here, we demonstrate the PTE response can be reshaped when Ca$_3$Ru$_2$O$_7$ undergoes meta-magnetic phase (MMP) transition driven by both temperature and magnetic field. Firstly, a sign change is observed crossing T$_S$ = 48 K and the linear polarization angle dependent PTE current maximizes along a-axis above T$_S$ while maximizes along b-axis below T$_S$, which indicates that the antiferromagnetic spin order contributes to such spatial anisotropy. Secondly, in the temperature range of around 40 ~ 50 K, the PTE current is found to be sharply suppressed when external magnetic field is applied in plane along a-axis but is only gradually suppressed when applied field is along b-axis which gives out two critical fields. We attribute such suppression of PTE current**

under magnetic field to the suppression of the spin entropy in the phase transition between the antiferromagnetic state and the MMP state and the H-T phase diagrams of $Ca_3Ru_2O_7$ is redrawn accordingly. Compared to previously work which trying to understand the magnetic phase transition in $Ca_3Ru_2O_7$, such as neutron scattering, specific heat, and other advanced transport measurements, our work provides a more convenient yet efficient method, which may also find applications in other correlated spin materials in general.

## 1. Introduction

PTE effect has been exploited as a new method other than Photopyroelectric (PPE) effect to characterize the phase transition of materials since it can sense the thermal parameters variation during the phase transition. [1-2] Traditional PPE or recently developed PTE method uses normally three- or four-layer measurement configuration, [3-5] exploiting materials with well characterized and superior thermoelectric (TE) property as pyroelectric (PE) or TE sensor. Dynamic thermo parameters such as thermo diffusivity, effusivity, and specific heat can be retrieved to portrait the phase transition. For instance, the antiferro-paramagnetic phase transition of $Cr_2O_3$ was successfully studied by PTE method by using Liquid TE sensor.[6] The LTE sensor is located between the metal electrode and the sample. A thin layer of silver is deposited on the backside of the sample, forming the second electrode for the PTE signal. Hence the thermal properties in the sample can be calculated given that thermoelectric coefficient of the LTE sensor is known. However, neither PPE or PTE method has simple and convenient measurement configuration since they both use external PE/TE sensors. One natural question is raised that weather PTE measurement can characterize phase transition without the assistance of external sensors. In this work, we demonstrate that the simplified PTE method can characterize the magnetic order and MMP in $Ca_3Ru_2O_7$, which is a well investigated perovskite oxide material.[7] The actually measured photocurrent is mainly from PTE effect and it turns out to be wideband, zero-bias driven and stable. By combining cryogenic and magnetic field dependent PTE measurement, the sharp PTE current variation of $Ca_3Ru_2O_7$ has been discovered when $Ca_3Ru_2O_7$ undergoes phase transition from AFM-a to AFM-b state at ~ 48 K. The application of in-plane magnetic field along b-axis causes the MMP transition and leads to the drastic suppression of the PTE current which we attributed to suppression of the spin entropy contribution. Although our simplified PTE method cannot measure the thermal parameters directly, our work provides an alternative and more convenient way to characterize the phase transitions related to correlated electronic or spin materials.

Layered ruthenates in the Ruddlesden-Popper series family $(Sr, Ca)_{n+1}Ru_nO_{3n+1}$ as a canonical complex transition metal oxide system has attracted great attention.[8-10] It has rich interactions among charges, spins, orbits, and crystal lattices resulting in a wealth of physical phenomena, including superconductors, ferromagnets, antiferromagnetic metals, structural phase transitions, magnetic phase transitions, etc.[11-16] In the Ruddlesden-Popper family $(Sr, Ca)_{n+1}Ru_nO_{3n+1}$ as the reduction of cation radius, system change from the quantum magnet $Sr_3Ru_2O_7$ to antiferromagnetic metal $Ca_3Ru_2O_7$ ;[17-19] as well as the increase in the number of perovskite $RuO_2$ layers that leads to the transition from a band-dependent Mott insulator $Ca_2RuO_4$ to the metallic $Ca_3Ru_2O_7$ with a k-dependent gap.[20-21] Ruddlesden-Popper $Ca_3Ru_2O_7$ has a rich phase diagram. $Ca_3Ru_2O_7$ is a paramagnetic (PM) metal at room temperature. [22] At $T_N$ = 56 K, $Ca_3Ru_2O_7$ exhibits a transition from PM order to antiferromagnetic order (AFM-a), which is characterized by the double layer is antiferromagnetically coupled along c-axis. The ferromagnetic perovskite bilayer with a‑axis, which is the short axis of the crystal. [14] At $T_S$ = 48 K, the second magnetic transition occurs, from AFM-a to AFM-b; the easy axis switches from a-axis to b-axis. [23-24] The transition at $T_S$ is also accompanied by a sharp increase in in-plane resistivity. Angle-resolved photoemission measurements reveal the destruction of a large hole-like Fermi surface upon cooling at 48 K.[9,16] With the appearance of an insulating-like pseudogap, $Ca_3Ru_2O_7$ undergoes semimetal-semiconductor transition.[25] Coincident with $T_S$, $Ca_3Ru_2O_7$ undergoes a first-order isosymmetric transition marked by a discontinuous change in the lattice parameters: the c-axis lattice constant is shortened, while those of the a- and b-axis are enlarged.[26] At AFM-b phase, a higher field transition under the magnetic field along the b-axis is a transition into a canted antiferromagnetic (CAFM) phase in which the magnetic moments on Ru ions are partially polarized along the direction of the applied field.[27] In the temperature range of 40 K to 48 K, the lower boundary of the magnetic field strength required for MMP phase transition gradually decreases with temperature, from 5.23 T (40 K) to 1.95 T (48 K).[23]

## 2. Results and Discussions

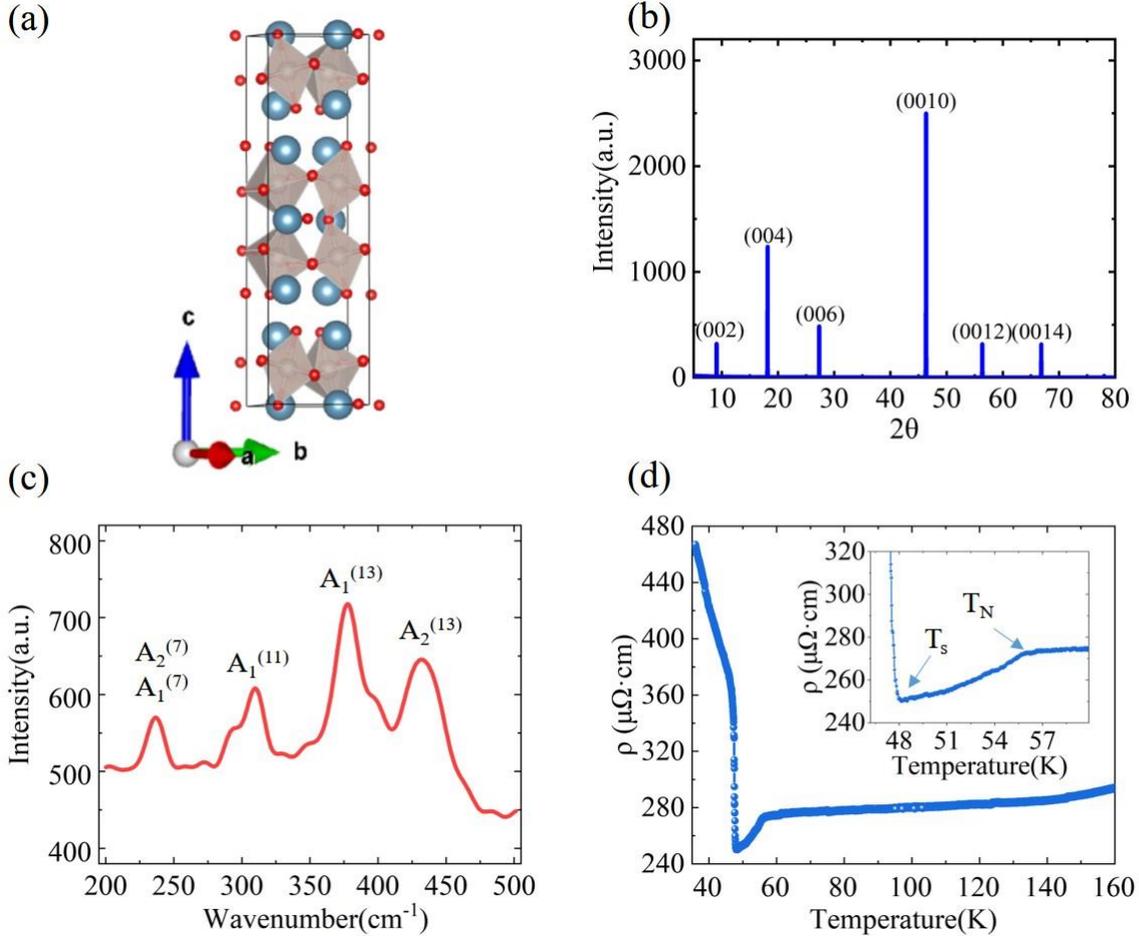

**Figure 1.** Basic characteristics of $Ca_3Ru_2O_7$ single crystal(a) Crystal structure of $Ca_3Ru_2O_7$. The blue spheres represent Ca atoms, the red spheres represent O atoms and the grey spheres represent Ru atoms. (b) X-ray diffraction pattern of $Ca_3Ru_2O_7$ crystals. (c) Raman spectra of $Ca_3Ru_2O_7$ crystals. (d) The temperature dependence of resistivity of $Ca_3Ru_2O_7$ crystals. The inset figure shows $T_N$ = 56 K and $T_S$ = 48 K.

The $Ca_3Ru_2O_7$ samples were prepared using the floating-zone (FZ) method. Optical micrograph of a typical $Ca_3Ru_2O_7$ crystal sample shows a clean and smooth surface obtained by mechanical cleavage. Due to its metallic nature, the crystal appears shiny black under ordinary white light. The sample is 1 mm * 1 mm * 0.3 mm single domain crystal. **Figure 1**a shows the crystal structure of $Ca_3Ru_2O_7$. $Ca_3Ru_2O_7$ is composed of perovskite double layer $(CaRuO_3)_2$ and $RuO_6$ octahedrons, separated by an additional CaO rock salt layer. The small ion size of Ca causes the large coupling rotation and tilt of the $RuO_6$ octahedron that constitutes the perovskite-like member of the structure, resulting in a non-centrosymmetric crystal structure. The crystal structure has the orthorhombic symmetry with $Bb2_1m$ space group (No.36($C_{2v}$), lattice parameters are a = 5.37 Å, b = 5.52 Å, c = 19.53 Å)[26] with rotation and tilting of $RuO_6$ octahedra. It has the magnetic moment ferromagnetically aligned within the $RuO_2$ bilayers and antiferromagnetically aligned between the bilayers. The x-ray diffraction patterns in Figure 1b show that $Ca_3Ru_2O_7$ crystal possess pure bilayer phase. As shown in Figure 1c, the Raman spectrum of $Ca_3Ru_2O_7$ has been measured in the range of 200-500 cm$^{-1}$ at room temperature. Four characteristic peaks of $Ca_3Ru_2O_7$ can be observed at 244, 311, 386 and 435 cm$^{-1}$. The peak centered at 435 cm$^{-1}$ can be assigned to the $A_2$ phonon mode, while the other three Raman peaks can be attributed to the $A_1$ phonon mode. All the observed Raman peaks are well in line with the phonon assignment reported earlier. [28] This indicates that high-quality $Ca_3Ru_2O_7$ crystals are obtained using the FZ growth method. Figure 1d shows the temperature dependence of resistivity. At high temperatures, it shows metallic behavior (the resistivity decreases as the temperature decreases, show that the $Ca_3Ru_2O_7$ is in the PM state. At $T_N$ = 56 K the resistivity drops sharply as the temperature continues to drop, which corresponds to the beginning of the AFM transition. While at $T_S$ = 48 K, the resistivity increases as the temperature decreases indicating the second magnetic transition occurs, from AFM-a to AFM-b. The transition at $T_N$ and $T_S$ agrees with the previous report. [29]

The photocurrent response is measured on the bulk $Ca_3Ru_2O_7$ single crystal fabricated device on silicon substrate with 300 nm Oxide layer. As shown in **Figure 2**a, compared to previous four layered device of PTE method, our PTE device only has sample layer with only normal contacts measuring the photocurrent, which demonstrates the simplicity of our method. The I-V characteristic curve under dark (dark line) and steady-state 532 nm light in the air at room temperature is shown in Figure 2b. The positive and negative denote the reverse of the defined polarity of the electrodes. The red and blue lines show a photocurrent of about 5 μA under zero voltage with respect to the dark I–V curve when the two sides of the sample were exposed to the laser beam with a power of approximately 20 mW. This photocurrent is almost unaffected by bias voltage, so subsequent experiments are all measured under zero bias. As shown in inset of Figure 2b, the photocurrents reach two maxima and has opposite signs when the beam illuminated the two sides ($Ca_3Ru_2O_7$–metal contacts), with no photocurrents induce when the beams focus on the center of channel, which are typical PTE effect.[30-31] The PTE effect arises when the laser causes local heating, generating a temperature difference between the two electrodes. Subsequently, the carriers diffuse from the hot region to the cold region, resulting in a photocurrent. Figure 2c shows the PTE current response over time with 10mW incident power under illumination by different lasers (375 nm, 473 nm, 532 nm, 635 nm, 1064 nm, and 4.7 μm). It shows typical photo-switching behavior in all excitation wavelength from UV to MIR at room temperature. The ultrabroadband photosensitive properties indicates that the response is PTE current response and the $Ca_3Ru_2O_7$ crystals can be used as active materials to fabricate ultrabroadband photodetectors. It also helps to select the most sensitive wavelength for subsequent experiments. Figure 2d displays the power dependence of the PTE current. All these light sources show almost linear behavior in a considerable power range, indicating that the PTE response based on $Ca_3Ru_2O_7$ can effectively distinguish different incident light intensities.

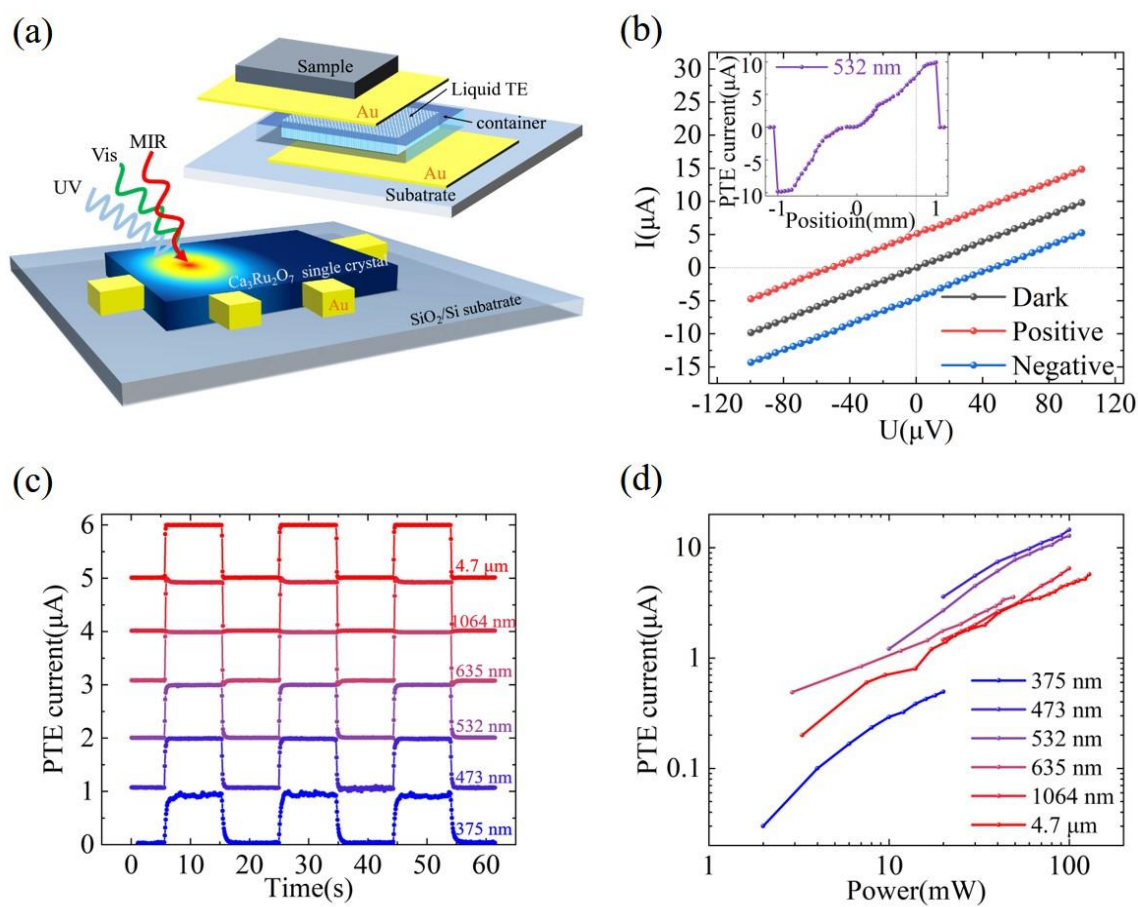

**Figure 2.** Simplified PTE effect of $Ca_3Ru_2O_7$ (a) Schematic plot for comparation of previous PTE method and our simplified version. (b) Current–voltage (I–V) curves of the $Ca_3Ru_2O_7$-based detector without illuminations (dark line) and with laser illumination of λ = 532 nm at the positive (blue line) and negative (red line) side. Inset shows position dependent PTE response. (c) PTE responses of the $Ca_3Ru_2O_7$-based detector under illumination by different lasers. (d) Power dependence of the PTE current under the aforementioned wavelengths.

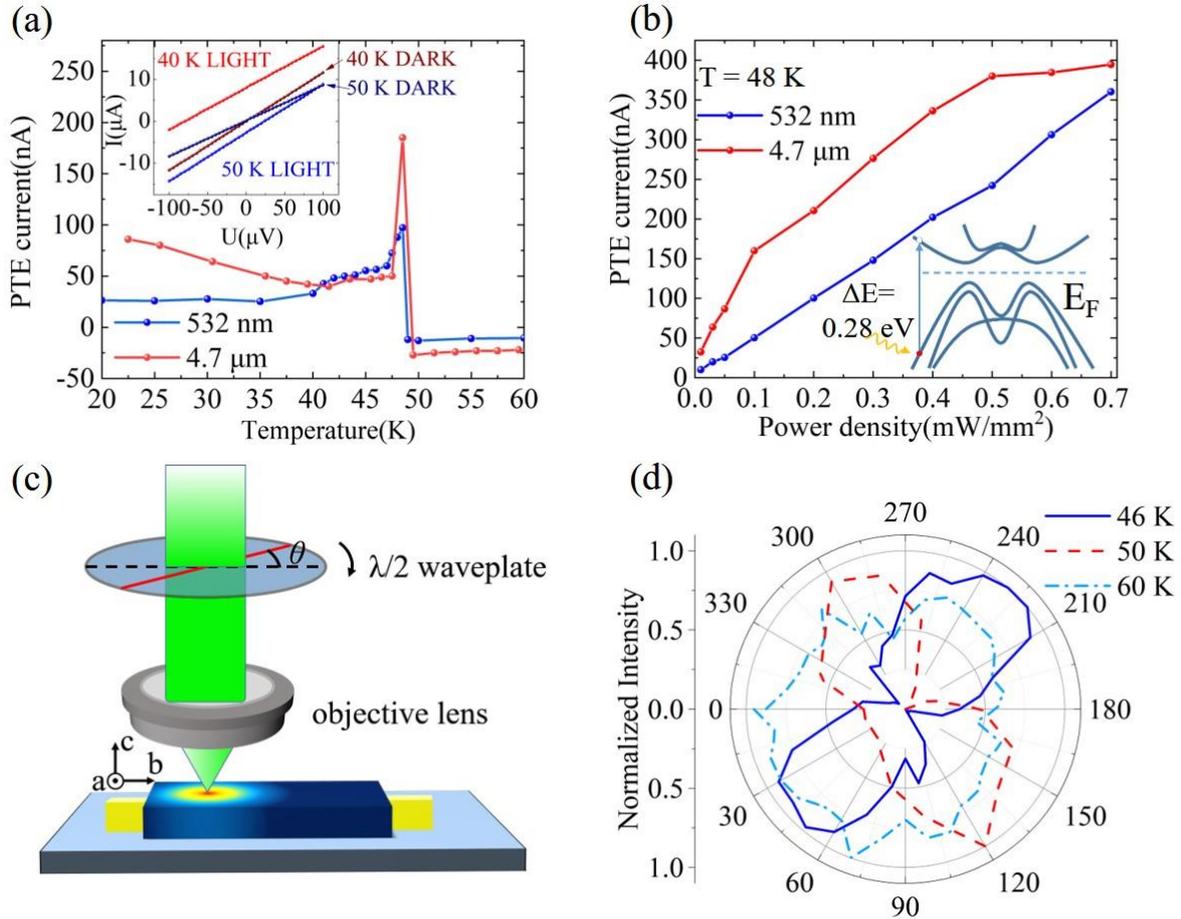

**Figure 3.** PTE response variation crossing $T_S \sim 48$ K (a) PTE response changes with temperature under laser illumination of $\lambda = 532$ nm(blue) and $\lambda = 4.7$ μm (red). Inset shows the I–V curves without illuminations and with laser illumination of $\lambda = 532$ nm at 40 K and 50 K. (b) The current response as the function of light power at illumination of $\lambda = 4.7$ μm at 48 K. Inset shows the electronic band structures at 48 K. (c) Schematic plot of the test setup for the angle dependent PTE response. (d) The angle dependence of PTE response at 46 K, 50 K and 60 K.

Previous work has revealed MMP transition at $T_S = 48$ K in $Ca_3Ru_2O_7$ by sophisticated transport measurements or neutron diffraction spectroscopy. However, those methods require complex techniques. Our PTE method is a simple one and can be easily combined with cryogenic and magnetic field environment. We expect our simplified PTE method is also efficient to characterize MMP. We studied the PTE response of $Ca_3Ru_2O_7$ devices at 30-60 K, paying particular attention to MMP transition region (around $T_S$ = 48 K) and found two characteristics. Firstly, as shown in Figure 3a, when temperature cools from 50 K to 40 K, the photocurrent has a sign change and a significant increasement in its amplitude under 0.2 mW power. From the I-V curve shown in the inset of Figure 3a, one can see at 40 K, photocurrent is linearly bias voltage dependent, which indicates that the semiconducting like behavior. This is consistent to previous finding that the Seebeck coefficient undergoes a sharp sign change across the phase transition and a small pseudogap opening.[9] Secondly, the PTE current peaks at $T_S$ where the value of 4.7 μm light is higher than that of 532 nm light as shown in Figure 3a. Previous band-structure calculations based on density-functional theory (DFT) shows that:[16,32] At $T_S$ < T< $T_N$ the electron and hole bands at the zone center develop a strong hybridization with multiple bands crossing $E_F$, this band structure is nearly semi-metallic. At $T_S$ = 48 K a pseudo-gap opens at $E_F$, and total density of states (DOS) of $Ca_3Ru_2O_7$ shows that the band is unoccupied near $E_F$ which mediated by spin–orbit coupling. A semimetal-semiconductor transition occurs as a pseudo-gap opens at $E_F$ in MMP transition. When the band is nearly full occupied (in AFM-a region), the probability of absorbing a photon to produce PTE response is reduced compared with that the band is unoccupied near $E_F$ when pseudo-gap exists (in MMP transition region). This was also seen in optical spectroscopy measurements upon cooling through $T_S$ = 48 K.[25] One can see from the inset of Figure 3b, the electrons can be photoexcited to contribute to PTE current directly without complicated electron phonon scattering for light of 4.7 μm since its photon energy matches the band energy difference while the visible light does not match, therefore, the PTE responsivity rate vs light power for 4.7 μm light is higher than the visible light as shown in Figure 3b. As shown in Figure 3c, a half-wave plate is used to rotate the normally incident light polarization to realize angle dependent PTE response measurement. In Figure 3d, the normalized photocurrent dependence on the azimuth angle *θ* of the incident linear polarized laser is plotted for different temperatures. An apparent anisotropy pattern emerges on the photocurrent *PTE(θ)*, revealing an overall two-fold symmetry. This two-fold symmetry is only observable for T<56 K, which indicate its magnetic origin. Most notably, a 90° shift is found between the *PTE(θ)* pattern at 46 K and 50 K. The major difference of the magnetic state at these two temperatures, as discussed before, was the AFM easy axis (a-axis for the high-T AFM-a state, b-axis for the low-T AFM-b state). Now the *PTE(θ)* reaches its maxima when the electric field of the incident photon aligns along the easy axis in both AFM-a and AFM-b states (through assistence of polarized optical reflection method.)[33]This magnetization-dependent photocurrent carries important insight. In view of the entropic origin of the thermoelectricity, a plausible mechanism is that the enhancement of *PTE(θ)* consists the contribution of spin entropy.[34-37] While at present it is difficult to seek for a quantitative assessment of the validity of the spin entropy contribution and its magnitude, especially in an partially itinerant *d*-electron system, the observed *PTE(θ)* patterns provide an tempting direction for further exploration and perhaps a feasible experimental probe for such mechanisms, which are typically elusive for other probes. Nevertheless, such photocurrent enhancement at the spin alignment direction in antiferromagnetic materials may be alternatively explained by the optical linear birefrangence, which the absortion of linearly polarized light maximizes at the direction of the antiferromagnetic spin alignment direction,[38-39] and hence the PTE current correspondingly demonstrates such linear polarization dependence.[40]

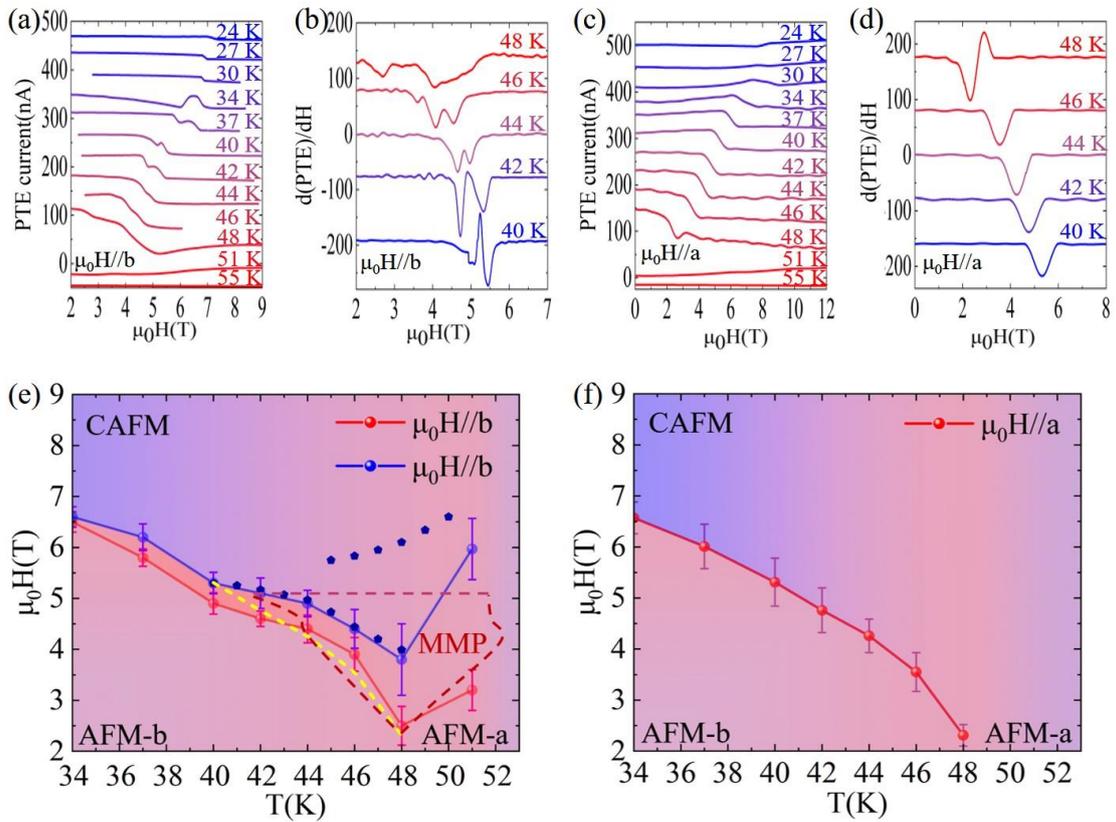

**Figure 4.** Magnetic field dependence of PTE response in $Ca_3Ru_2O_7$ (a-d) PTE response at various temperatures under magnetic field along the b-axis(a&b) and a-axis(c&b). (e-f) Phase diagrams of $Ca_3Ru_2O_7$ under the magnetic field along the b-axis(e) and a-axis (f), the red and blue circle dots represent the peak positions derived from (b&d). The dashed yellow line represents the data of neutron scattering and the blue pentagon dots represent the data from Hall measurement in (e), which are reproduced with permission. [23] 2019, Nature publishing group.

**Figure 4**a&c shows the PTE response with the applied in-plane magnetic field at different temperatures. The main feature is the significant suppression of the PTE current at certain magnetic fields, which is only observed at 48 K and below. However, the suppresion behavior is in stark contrast when magnetic field is along a-axis and b-axis. For instance, when magnetic field along a-axis, there is only one sharp decrease appear in the PTE current curves. When magnetic field along b-axis, there are plateaus appear in the PTE current decrease process. Taking $T_S = 48$ K as an example, the PTE current starts to decrease at ~ 2.5 T, the whole process ends until about 5 T, while have a wide plateau in between. As the temperature lowered below $T_S$, the magnetic field required to produce the PTE current suppression increases. By taking the differentiate point of d(PTE)/dH, we derive the turning points at different temperatures for magnetic field in both direction, as displayed in Figure 4b&d. These critical points can redraw the phase diagram partly as demonstrate in Figure 4e&f. For comparism, we replot the small-angle neutron scattering (SANS) experiments result by the light yellow dashed line, and the data of hall measurement result by dark blue shapes from previous work.[23] It can be found that our measurement results fit the two curves in the figure very well, which means that we can use the measurement of the PTE response to characterize the magnetic order phase transition.

We now turn to discuss why PTE current is suppressed significantly under external in plane magnetic field. This can also be probably explained with the scenary of spin entropy contribution mentioned above. While the magnetization is reported to be small at AFM-b state, it is dramatically increased and saturated at much higher value at the critical magnetic fields which draw the phase boundary between CAFM and AFM-b phase.[23] The increasement of magnetization means the alignment of spins, equally, the charge carriers lose their spin entropy. This phenomenon is only obvious at the temperature region of AFM-b state which has localized carriers.[10] The magnetic field applied in either a-axis or b-axis direction tends to switch from AFM-b state to a canted AFM state, where the spin entropy of localized carriers in AFM-b will be lost during such switch process and hence the PTE current suppression. However, compared to the simple switch from AFM-b state to a canted AFM-a state under magetic field along a-axis, the switch process from AFM-b to canted AFM-b is more intricate, accomanying with a meta magnetic transition in between.[23] Our PTE current suppression process has an intermediate plateau when external magnetic field is along b-axis seems to be consistent with such scenaro.

## 3. Conclusion

In summary, we have constructed simple device of $Ca_3Ru_2O_7$ single crystal, and measured its PTE response as a function of temperature and magnetic field. We find that the value of the PTE response has changed nearly ten times under laser illumination of $\lambda = 532$ nm and $\lambda = 4.7$ μm, at temperature $T_S$. The change in the PTE response coincides with the change in the energy band structure caused by the SOC in the MMP transition or alternatively, the contribution from spin entropy of the localized carriers. Applying magnetic field in different directions changes the magnetic order and suppresses the PTE response. Compared with traditional thermodynamic transport measurement, our simplified PTE method complements and confirms each other, and the difference between the two may be caused by different physics. Our simplified PTE method is convenient yet efficient while maintaining the measurement accuracy of phases evolution, which has great potential in investigation of other correlated electronic or spin materials.

## 4. Experimental Section/Methods

*Device Fabrication*:

The photo-response based on bulk $Ca_3Ru_2O_7$ were fabricated using 1 mm*1 mm*0.3 mm single domain $Ca_3Ru_2O_7$ crystal on the Silicon substrate with 300 nm Oxide layer. Thin gold wires were connected to freshly cleaved surface by covering the silver paint completely to enable a good Ohmic contact.

*PTE response Measurements*:

All experimental results were performed under zero bias, unless otherwise mentioned. The PTE response performance was tested with different wavelength laser, including UV (375 nm), visible (473 nm, 532 nm, 635 nm), near-infrared (1064 nm) and mid-infrared (4.7 μm). At $\lambda =$ 375,473,532,635,1064 nm excitation light sources are continuous-wave solid-state lasers (Changchun New Industries Optoelectronics Technology Ltd.), and at mid-IR 4.7 μm light source is continuum wave quantum cascade laser. The current–voltage (I–V) characteristics of the devices were measured using a Keithley 2450 sourcemeter loaded in a closed cycle optical cryostat (SHI-4-1, Janis Ltd.) form room to low temperature. Magnetic PTE response measurements were performed in a closed cycle cryostat (Oxford Instruments Inc.). Low noise preamplifier (SR 570) and lock-in amplifier (SR 830) were employed to measure the PTE response.

**Supporting Information**

Supporting Information is available from the Wiley Online Library or from the author.

**Acknowledgements**

L.J Li acknowledges the funding from National Key R&D Program of China (2019YFA0308602), National Science Foundation of China (11774308&12174336) and the Zhejiang Provincial Natural Science Foundation of China (LR20A040002).

Received: ((will be filled in by the editorial staff))
Revised: ((will be filled in by the editorial staff))
Published online: ((will be filled in by the editorial staff))

The Photo-Thermo-Electric response is measured on the $Ca_3Ru_2O_7$ single crystal device. Significant modulation in the PTE response is attributed to the contribution from spin entropy of the localized carriers in certain magnetic ordering state. The critical points of d(PTE)/dH at different temperatures reconciles well with boundary of phase diagram drawn by previous research.

Qiang Chen[1], Jialin Li[1], Huanfeng Zhu[1], Tian Zhang[1], Wei Tang[1], Hui Xing[2], Jin Peng[3], Zhiqiang Mao[4], Linjun Li[1*]

**A simplified method characterizing magnetic ordering modulated photo-thermoelectric response in noncentrosymmetric semimetal $Ca_3Ru_2O_7$**

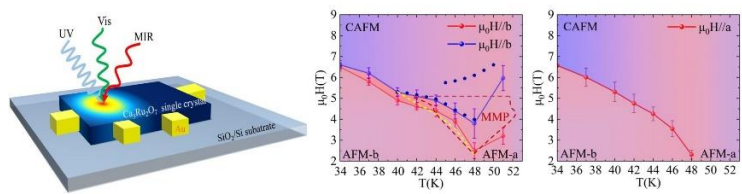